\documentclass[journal,9pt,twoside,web]{ieeecolor}
\usepackage{generic}

\usepackage{ifpdf}
\ifpdf
\fi

\usepackage[T1]{fontenc}
\usepackage[utf8]{inputenc} 
\usepackage{graphicx,color} 
\usepackage{times} 
\usepackage{amsmath} 
\usepackage{amssymb}  
\usepackage{caption}
\usepackage{subcaption}
\usepackage{stackrel}
\usepackage{tcolorbox}

\usepackage{url}
\usepackage{multirow}
\usepackage{cite}

\usepackage{algorithm}
\usepackage[noend]{algpseudocode}
\usepackage{pgf}
\usepackage{pgfplots}
\usepgfplotslibrary{fillbetween}
\usepackage{tikz}
\usetikzlibrary{arrows,automata}
\usetikzlibrary{intersections}
\usetikzlibrary{calc}
\usetikzlibrary{shapes}
\usetikzlibrary{positioning}
\usetikzlibrary{decorations.text}
\pgfdeclarelayer{background}
\pgfdeclarelayer{foreground}
\pgfsetlayers{background,main,foreground}
\usetikzlibrary{calc,patterns,decorations.pathmorphing,decorations.markings}
\usepackage{dsfont}
\usepackage{booktabs}

\let\labelindent\relax
\usepackage{enumitem}
\renewcommand{\theenumi}{\arabic{enumi}}

\renewcommand{\theenumii}{\arabic{enumii}}

\renewcommand{\theenumiii}{\arabic{enumiii}}

\renewcommand{\theenumiv}{\arabic{enumiv}}

\newcommand{\PP}{{\mathcal P}}

\newcommand{\XX}{{\mathcal X}}
\newcommand{\Aa}{\mathcal A}
\newcommand{\CC}{{\mathcal C}}

\newtheorem{definition}{Definition}

\newtheorem{proposition}{Proposition}
\newcommand{\rr}{\mathbb{R}}
\newcommand{\st}{\mbox{s.t.~} }

\newcommand{\beqar}{\begin{eqnarray}}
\newcommand{\eeqar}{\end{eqnarray}}
\newcommand{\beqarno}{\begin{eqnarray*}}
\newcommand{\eeqarno}{\end{eqnarray*}}
\newcommand{\ba}[1]{\begin{array}{#1}}
\newcommand{\ea}{\end{array}}
\newcommand{\diag}{\mathop{\rm diag}\nolimits}

\newcommand{\rank}{\mathop{\rm rank}\nolimits}

\newcommand{\smallmat}[1]{\left[ \begin{smallmatrix}#1 \end{smallmatrix} \right]}

\usepackage{xcolor}
\definecolor{change_color}{RGB}{204, 102, 0}
\newcommand{\change}[1]{{\color{black}#1}}

\newtheorem{runex}{Example}

\definecolor{mycol}{RGB}{19,48,128}
\usepackage{hyperref} 
\hypersetup{
	colorlinks = true,
	citecolor = mycol,
	linkcolor = mycol,,
	urlcolor = black
}

\graphicspath{{figs/}}

\begin{document}

	\title{\LARGE Solving Multiparametric Generalized Nash Equilibrium Problems\\ and Explicit Game-Theoretic Model Predictive Control}

    \author{Sophie Hall, Alberto Bemporad, \IEEEmembership{Fellow, IEEE} 
	\thanks{Sophie Hall is with the Automatic Control Laboratory, ETH Z\"urich, Switzerland (e-mail: \texttt{\scriptsize shall@control.ee.ethz.ch}).\\
    Alberto Bemporad is with the IMT School for Advanced Studies Lucca,
    55100 Lucca, Italy (e-mail: \texttt{\scriptsize alberto.bemporad@imtlucca.it}).\\
    This work was supported by the Swiss National Science Foundation under the NCCR Automation (grant 51NF40 225155) and by the European Union (ERC Advanced Research Grant COMPACT, No. 101141351). Views and opinions expressed are however those of the authors only and do not necessarily reflect those of the European Union or the European Research Council. Neither the European Union nor the granting authority can be held responsible for them.
    }}
	
\maketitle
\thispagestyle{empty}
	
\begin{abstract} 
We present a method for computing explicit solutions to parametric generalized Nash equilibrium (GNE) problems with convex quadratic cost functions and linear coupling and local constraints. Assuming that the parameters enter only the linear terms of the cost functions and the constraint right-hand sides, we provide the exact multiparametric solution of the GNE problem. Such a solution enables: ($i$) minimal real-time computation; ($ii$) inherent interpretability and explainability, as well as exact enumeration of all multiple equilibria; ($iii$) selection of desired GNE solution types in the case of infinitely many equilibria; and ($iv$) zero-shot updates of the GNE solution in response to changes in constraint right-hand sides and/or linear costs. In line with explicit model predictive control (MPC) approaches, we apply our method to solve game-theoretic MPC problems, also known as receding horizon games, explicitly. We compare its performance against centralized solvers in a battery charging game and a toy two-mass-spring-damper system control problem. A Python implementation of the algorithms presented in this paper is available at \url{https://github.com/bemporad/nash_mpqp}.
\end{abstract}

\begin{keywords}
Generalized Nash equilibrium problems, multiparametric programming, game-theoretic MPC
\end{keywords}

\section{Introduction}

Generalized Nash equilibria (GNEs) model strategic interactions among self-interested agents whose objectives and constraints are coupled~\cite{facchinei2009generalized}. They have received increasing attention in multi-agent control, as their strategic structure helps overcome issues with fixed control policies in which other agents' actions are treated as constraints or disturbances. A notable example is the frozen robot problem in highway merging of autonomous vehicles, resolved through a GNE policy~\cite{lecleach2022algames}. GNEs are used in broad classes of applications, including supply chains~\cite{hall2024receding}, energy~\cite{atzeni2012demand, hall2022receding}, transportation~\cite{bassanini2002allocation}, and telecommunications~\cite{pavel2012game}.


On the algorithmic side, various approaches exist for solving GNEs, including: ($i$) operator splitting methods~\cite{belgioioso2022distributed, yi2019operator}; ($ii$) augmented Lagrangian methods~\cite{pang2005quasi, lecleach2022algames}; ($iii$) interior-point methods~\cite{dreves2011globally}; ($iv$) Newton methods~\cite{facchinei2007generalized, facchinei2003finite, dreves2011globally, liu2024input}; and ($v$) learning-based methods~\cite{FB25}. In recent years, a major focus has been on algorithms for variational GNE (v-GNE) problems, which impose homogeneous Lagrange multipliers across agents and can be reformulated as variational inequalities~\cite{facchinei2009generalized}, for which both semi-decentralized~\cite{belgioioso2023semi} and fully distributed~\cite{yi2019operator, bianchi2022fast} methods are available.
Yet all these methods face an inherent trade-off: in a centralized setting, they suffer from a high computational burden increasing with the number of agents, whereas in a distributed setting, they require many communication rounds~\cite{lecleach2022algames, yi2019operator}. Most importantly, to date, no algorithm computes the entire (and possibly infinite) set of GNE solutions in the multidimensional setting, nor reliably selects and computes GNEs that are not variational solutions.

In this paper, we propose a method to compute explicit GNE solutions as functions of one or more parameters. The idea builds on multiparametric quadratic programming (mpQP) solvers~\cite{MR64,Fia83,bemporad2002explicit,tondel2003algorithm,GBN11,arnstroem2024high,schaller2025automatic} and their use in implementing constrained linear model predictive controllers without online optimization~\cite{bemporad2002explicit}.
In the game-theoretic setting, explicit GNE solutions inherit the same advantages as explicit mpQP solutions: ($i$) inexpensive real-time evaluation via lookup tables; and ($ii$) interpretability, since solutions can be analyzed before deployment and any desired equilibrium selected a priori, unlike existing non-parametric methods that only return a specific GNE. In the multi-agent setting, two additional advantages emerge: ($iii$) minimal communication, as only parameters are shared among agents with no iterative inter-agent rounds; and ($iv$) privacy preservation, since no sensitive data such as state or input trajectories are exchanged at runtime. Thus, explicit GNE solutions offer a pathway to make this solution concept more practical and deployable in real-world applications. In particular, due to low computational and communication requirements, we expect that our results will enable applying the GNE solution concept in practical settings characterized by fast actuation rates, low on-board computation resources, and communication limitations. 

This paper makes two main contributions. First, we develop an algorithm to compute explicit GNE solutions in the linear-quadratic setting, without any restriction on the monotonicity properties of the game nor on the variational nature of the equilibria. Second, we present approaches to selecting GNE solutions in the presence of multiple equilibria, which has not been possible with existing methods. A running example provides insight into each step and its interpretation. We also demonstrate applicability in a game-theoretic MPC setting, providing simulation results for a battery charging competition with coupled grid load limits and for a toy mechanical system control problem.

The remainder of this paper is organized as follows. Section~\ref{sec:MultiparametricGNEs} formulates the multiparametric GNE problem and presents the algorithmic procedure used to compute explicit GNE policies. Section~\ref{sec:GameTheoreticMPC} introduces a game-theoretic MPC problem and presents simulation examples. Section~\ref{sec:Conclusion} concludes the paper.

 \section{Multiparametric Generalized Linear Quadratic Games}\label{sec:MultiparametricGNEs}
 
Consider a parametric generalized Nash equilibrium problem with $n_x$ decision variables
in which $N$ agents take decisions individually, each one responding with a decision \change{$\bar x_i(x_{-i},p)$} to the other agents' decisions $x_{-i}$, given a parameter vector $p$ contained in a set $\PP\subseteq\rr^{n_p}$. Each agent aims to minimize their own quadratic cost function, which is strictly convex and subject to linear constraints, as follows:
\begin{subequations}
\beqar
\change{\bar x_i(x_{-i},p)}\!&=&\!\arg\min_{x_i} \frac{1}{2}x^\top Q_i x+(c_i+F_ip)^\top x \label{eq:mpQPi-cost}\\
                  & &\st      Ax\leq b + Sp\label{eq:mpQPi-constraints-ineq}
\eeqar
\label{eq:mpQPi}%
\end{subequations}
where $x_i$ is the decision vector of agent $i$, $x_i\in\rr^{n_i}$, $\sum_{i=1}^Nn_i=n_x$, $x_{-i}\in\rr^{n_x-n_i}$ is the vector collecting all agents' decisions $x_j$, $j=1,\ldots,N$, $j\neq i$, and $Q_i = (Q_i)^\top \in\rr^{n_x\times n_x}$. We define as $Q_{ii}\succ0$ the diagonal block of $Q_i$ corresponding to the variable indices associated with agent $i$ and similarly the off-diagonal blocks as $Q_{-ii}$. Further, $c_i\in\rr^{n_x}$, $F_i\in\rr^{n_x\times n_p}$ define the cost function of agent $i$
given $x_{-i}$ and $A\in\rr^{n_A\times n_x}$, $b\in\rr^{n_A}$, $S\in\rr^{n_A\times n_p}$ define the linear inequality constraints\footnote{The approach described in this paper can be easily extended to include linear equality constraints $Ex=f+Wp$, where
$E\in\rr^{n_E\times n_x}$, $f\in\rr^{n_E}$, $W\in\rr^{n_E\times n_p}$.}. Possible local constraints on $x_i$ can be included in~\eqref{eq:mpQPi-constraints-ineq} 
by \change{embedding them into the} matrices $A$, $b$, $S$\change{, with the corresponding rows of $A$
containing nonzero entries only corresponding to agent $i$'s actions.}
Note also that, while each agent's problem~\eqref{eq:mpQPi} may depend only on a subvector $p_i$ of parameters in $p$, for maximum generality we have considered $p_i=p$, $\forall i=1,\ldots,N$, possibly zeroing suitable entries in $F_i$, $S$ 
to remove the dependence on unwanted parameters. 

The Karush-Kuhn-Tucker (KKT) optimality conditions for each agent's problem~\eqref{eq:mpQPi} are:
\begin{subequations}
\beqar
&&Q_{ii} x_i + Q_{-ii} x_{-i}+ \bar c_i+\bar F_i\left[\begin{smallmatrix}x_{-i}\\ p \end{smallmatrix}\right]+ A_i^\top\lambda_i = 0\\
&&A_i x_i \leq  b
 + \bar B_i \left[\begin{smallmatrix}x_{-i}\\ p \end{smallmatrix}\right]\\
&&\lambda_i \geq 0 \\
&&\lambda_i^\top(A_{i}x_i -b- \bar B_{i}\left[\begin{smallmatrix}x_{-i}\\ p \end{smallmatrix}\right]) = 0 \label{eq:CompSlack}
\eeqar
\label{eq:KKT}%
\end{subequations}
where $\bar c_i$ and $\bar F_i$ are collecting the linear terms of the cost function~\eqref{eq:mpQPi-cost} associated with $x_i$, 
$A_i$ as the submatrix of $A$ collecting the columns corresponding to the decision variables $x_i$ of agent $i$, and $\bar B_i$ as the matrix formed by appending $S$ to the submatrix of $-A$ obtained by collecting the columns corresponding to all other agents' decision variables $x_{-i}$.
We assume that all the functions and sets involved in~\eqref{eq:mpQPi} are known to a central entity, which is responsible for solving the multiparametric Nash equilibrium problem for all $p\in\PP$. In particular, our goal is to find an 
{\it explicit} solution $x^*(p)$ to the Nash equilibrium problem
\begin{equation}
x_i^*(p) = \bar x_i(x_{-i}^*(p),p),\ \forall i=1,\ldots,N,\ \forall p\in\PP
\label{eq:approximate_best_response}
\end{equation}
within a given hyper-box $\XX=\{x\in\rr^{n_x}:\ x_{\rm min}\leq x\leq x_{\rm max}\}$ \change{of interest. Throughout the paper, we denote by $x$ the stacked decision vector, by $x^*(\cdot)$ a parametric equilibrium, and by $\bar x(\cdot)$ a best-response map.}
Note that, in the case where lower (upper) bounds on $x$ are enforced in~\eqref{eq:mpQPi-constraints-ineq},
the values of $x_{\rm min}$ ($x_{\rm max}$) can be tightened accordingly. Further, if the projection on the $x$-space of the polyhedron defined by the constraints \eqref{eq:mpQPi-constraints-ineq} and $\mathcal{X}$ is bounded, then we have a solution for all $p$ in the projection on the $p$-space of the same polyhedron, see~\cite[Thm. 4.1]{facchinei2009generalized} for existence results of GNE problems.
\begin{runex}
To exemplify our developments, we consider the simple 2-agent GNEP discussed in the seminal paper~\cite{rosen1965existence}, which we turn into a multiparametric GNEP by introducing parameters $p_1$ in the cost of agent 1 and $p_c$ in the coupling constraint. The problems of the two agents are:

~\\
\begin{tabular}{c|c}
\begin{minipage}[c]{0.45\columnwidth}
~\\[-1em]
\[
\begin{aligned}
\displaystyle \min_{x_1\geq 0 } &  \frac{1}{2}(x_1)^2 - x_1x_2 + p_1\, x_1  \\
 \text{s.t.}& -x_1  -x_2 \leq p_{c}
\end{aligned}
\]
\end{minipage}
&
\begin{minipage}[c]{0.45\columnwidth}
~\\[-1em]
\[
\begin{aligned}
\displaystyle \min_{x_2\geq 0 } &  (x_2)^2 + x_1 x_2 \\
 \text{s.t.}& -x_1  -x_2 \leq p_c
\end{aligned}
\]
\end{minipage}
\\\hline
\begin{minipage}[c]{0.45\columnwidth}
\[
Q_1 = \left[\begin{smallmatrix}1 & -1\\ -1 &0\end{smallmatrix}\right],\; F_1 = \left[\begin{smallmatrix}0 & 1\\0 & 0\end{smallmatrix}\right] 
\]
\end{minipage}
&
\begin{minipage}[c]{0.45\columnwidth}
\[
Q_2 = \left[\begin{smallmatrix}0 & 1\\ 1 &2\end{smallmatrix}\right] 
\]
\end{minipage}
\end{tabular}
~\\
where $ p =[p_c\,,\,  p_1]^\top $ and with global constraint matrices 
\begin{align*}
A = \left[\begin{smallmatrix}
-1 & -1\\
-1 & 0\\
0 & -1
\end{smallmatrix}\right], 
\quad b = \mathbf{0}_{n_A}
\quad  S = \left[\begin{smallmatrix}
1 & 0\\
0 & 0\\
0 & 0
\end{smallmatrix}\right].
\end{align*}
To determine the critical regions of each agents' mpQP, we consider their active sets 
derived from the KKT conditions~\eqref{eq:KKT}:
~\\
\begin{tabular}{c|c}
\begin{minipage}[c]{0.45\columnwidth}
~\\[-1em]
\begin{align*}
\left[\begin{smallmatrix}
-1 \\
-1 \\
0 
\end{smallmatrix}\right] x_1 \leq  \left[\begin{smallmatrix}
x_2 + p_c\\
0\\
x_2
\end{smallmatrix}\right] \\[2ex] 
\left[\begin{smallmatrix}
\lambda_{1,1} \\
\lambda_{1,2}\\
\lambda_{1,3}
\end{smallmatrix}\right] \geq \left[\begin{smallmatrix}
0\\
0\\
0
\end{smallmatrix}\right]  \\[2ex] 
 x_1 -x_2 + p_1 - 
\lambda_{1,1} 
-\lambda_{1,2}
 = 0 \\[2ex] 
\lambda_{1,1}(-x_1-x_2 -p_c) = 0\\[2ex] 
\lambda_{1,2}(-x_1) = 0 
\end{align*}
\end{minipage}
&
\begin{minipage}[c]{0.45\columnwidth}
~\\[-1em]
\begin{align*} 
\left[\begin{smallmatrix}
-1 \\
0\\
-1 
\end{smallmatrix}\right] x_2\leq  \left[\begin{smallmatrix}
x_1 + p_c\\
x_1\\
0\\
\end{smallmatrix}\right] \\[2ex] 
\left[\begin{smallmatrix}
\lambda_{2,1} \\
\lambda_{2,2}\\
\lambda_{2,3}
\end{smallmatrix}\right] \geq \left[\begin{smallmatrix}
0\\
0\\
0
\end{smallmatrix}\right] \\[2ex] 
 2x_2 + x_1 - 
\lambda_{2,1} -
\lambda_{2,2}
  = 0 \\[2ex] 
\lambda_{2,1}(-x_2-x_1 -p_c) = 0\\[2ex] 
\lambda_{2,2}(-x_2) = 0
\end{align*}
\end{minipage}
\end{tabular}
~\flushright\QED
\end{runex}

Each solution \change{$\bar x_i(x_{-i},p)$} can be computed (in parallel, with respect to each $i$) by any multiparametric quadratic programming (mpQP) solver \change{which partitions the $(x_{-i},p)$-space into finitely many polyhedra, each associated with a fixed active set of constraints known as \emph{critical regions}~\cite{bemporad2002explicit}. As a consequence,  $\bar x_i(x_{-i},p)$ is a continuous and piecewise-affine (PWA) function of $(x_{-i},p)$:}
\begin{equation}
    \begin{aligned}
    \change{\bar x_i(x_{-i},p)} = &\left\{ \ba{ll}
    H_{-i}^1 x_{-i} + G_i^1 p + h_i^1& \mbox{if}\ (x_{-i},p)\in CR_i^1 \\
    \vdots & \vdots\\
    H_{-i}^{N_i} x_{-i} + G_i^{N_i} p + h_i^{N_i}& \mbox{if}\ (x_{-i},p)\in CR_i^{N_i} \\
    \ea\right.\\
    CR_i^j = & \left\{(x_{-i},p):\ C_{-i}^jx_{-i}+D_i^jp\leq e_i^j\right\}\\ &j=1,\ldots,N_i
    \end{aligned}
\label{eq:mpQP-i}
\end{equation}
where $N_i$ is the number of polyhedral regions defining the partition in the space of parameters $(x_{-i},p)$,
each one associated with a different active set $\Aa_j\subseteq\{1,\ldots,n_A\}$ of inequality constraints,
$N_i\leq 2^{n_A}$. \change{The matrices $H_{-i}^j$, $G_i^j$ and vector $h_i^j$ define the affine best response of agent $i$ within the polyhedral region $CR_i^j$ associated with the active set $\Aa_j$, which is in turn defined through the matrices $C_{-i}^j$, $D_i^j$ and vector $e_i^j$.}   
  
\subsection{Finding parametric equilibria}  
\label{sec:parametric-linear-system}  
For any given combination $\CC_k=(j_1,\ldots,j_N)$, $k\in\{1,\ldots,\prod_{i=1}^NN_i\}$, of affine best-response solutions and corresponding critical regions $CR_i^{j_i}$, determined by solving the mpQP's~\eqref{eq:mpQP-i}, $x$ is an equilibrium if and only if it satisfies the following parametric linear system
\begin{subequations}
\begin{equation}
    \left\{\ba{rcl}
    x_1 &=& H_{-1}^{j_1} x_{-1} + G_1^{j_1} p + h_1^{j_1}\\
    \vdots&&\vdots\\
    x_N &=& H_{-N}^{j_N} x_{-N} + G_N^{j_N} p + h_N^{j_N}
    \ea\right.
    \label{eq:mpQP-linear-system}
\end{equation}
and the linear inequalities
\begin{equation}
    C_{-i}^{j_i}x_{-i}+D_i^{j_i}p\leq e_i^{j_i}
    \label{eq:mpQP-intersection}
\end{equation}
\label{eq:mpQP-solution}%
\end{subequations}
are satisfied for all $i=1,\ldots,N$.

\begin{definition}[\change{GNEP critical regions}]
The polyhedron $CR_k$ given by the intersection of $\PP$ with the projection
of the polyhedron defined by~\eqref{eq:mpQP-linear-system}--\eqref{eq:mpQP-intersection} onto the $p$-space is called the \change{\emph{critical region of the GNEP}} associated with combination $\CC_k$ of affine best responses.
\label{def:CR_k}
\end{definition}
\change{As} we will detail later, we will provide techniques to characterize the critical regions $CR_k$ without necessarily computing the projections introduced in Definition~\ref{def:CR_k}, so that combinations $\CC_k$ leading to critical regions $CR_k$ with empty interiors can be discarded as well. Typically, this drastically limits the combinatorial explosion of the sought explicit solution due to considering all $\prod_{i=1}^NN_i$ combinations $\CC_k$. We remark that, differently from standard mpQP solutions, in the GNEP setting multiple critical regions $CR_k$ {\it may overlap in the $p$-space}, as different combinations $\CC_k$ may lead to different equilibrium solutions for the same $p$. In other words, we can exactly characterize all the equilibria that exists for a given $p$ by identifying the overlapping critical regions that contain it.

To characterize Nash equilibria $x^*$ as a function of $p$, let us first rewrite the linear system~\eqref{eq:mpQP-linear-system} in the compact form 
\begin{equation}
    M_x^k x=M_p^kp+M^k_1
    \label{eq:mpQP-linear-system-short}
\end{equation}
where $M_x^k\in\rr^{n_x\times n_x}$, $M_p^k\in\rr^{n_x\times n_p}$, $M_1^k\in\rr^{n_x}$ are the matrices and vector collecting the coefficients of the linear system~\eqref{eq:mpQP-linear-system} for the $k$-th combination $\CC_k$ of affine best responses.
We distinguish the following cases: $i$) $M_x^k$ is full rank $n_x$, $ii$) $M_x^k$ has rank $n_M<n_x$. 
From now on, we will drop the superscript $k$ to simplify our notation.

\change{
Note that~\eqref{eq:mpQP-linear-system} only imposes the equilibrium of the agents' decisions,
with no restriction on the dual variables associated with the coupling constraints. 
Hence,~\eqref{eq:mpQP-linear-system-short} characterizes all GNEs corresponding to $\CC_k$, not just the variational ones. We will discuss how to compute vGNEs in Section~\ref{sec:variational-GNE}.}

\subsection{Unique parametric Nash equilibrium}
When $M_x$ is full rank, and therefore invertible, 
the parametric Nash equilibrium is defined by the affine expression
\begin{subequations}
\begin{equation}
    x^*(p) = M_x^{-1}M_p p + M_x^{-1}M_1
\end{equation}
and the associated critical region $CR_k$ is the polyhedron
\begin{equation}
    CR_k=\{p\in\rr^{n_p}:\ C_{-i}^{j_i}x_{-i}^*(p)+D_i^{j_i}p\leq e_i^{j_i},\ i=1,\ldots,N\}.
    \label{eq:mpQP-solution-full-rank-CR}
\end{equation}
\label{eq:mpQP-solution-full-rank}%
\end{subequations}
The full dimensionality of $CR_k$ can be easily checked numerically,
e.g., by finding the Chebyshev radius of $CR_k$
via linear programming~\cite[Section 4.3.1]{BV04}.

\subsection{Infinitely many parametric Nash equilibria}
\label{sec:infinitely-many}
Consider now the case $\rank(M_x)=n_M<n_x$.
To characterize the solutions (if any exist) of~\eqref{eq:mpQP-linear-system-short}, we consider the singular value decomposition (SVD) of $M_x$:
\[
    M_x = U\diag(\sigma)V'
\]
where $U\in\rr^{n_x\times n_x}$, $V\in\rr^{n_x\times n_x}$ are orthogonal matrices and $\sigma\in\rr^{n_x}$ is a vector collecting the singular values of $M_x$ in decreasing order, $\sigma_i=0$ for $i=n_M+1,\ldots,n_x$.
Let us decompose $V=[V_1\ V_2]$, where $V_1\in\rr^{n_x\times n_M}$, $V_2\in\rr^{n_x\times (n_x-n_M)}$,
$\sigma=[\sigma_1'\ 0']'$, $\sigma_1\in\rr^{n_M}$, $U=[U_1\ U_2]$, where $U_1\in\rr^{n_x\times n_M}$, $U_2\in\rr^{n_x\times (n_x-n_M)}$ and consider the change of variables $y_1=V_1'x$, $y_2=V_2'x$, i.e., $y=V'x$. 
Since $x=V_1y_1+V_2y_2$, we can left-multiply both terms in the linear system~\eqref{eq:mpQP-linear-system-short}
by $U'$ which gives
\begin{equation}
    \begin{aligned}
    \diag(\sigma_1)y_1 &= U_1' M_pp+U_1' M_1\\
    0 &= U_2' M_pp+U_2' M_1.
    \end{aligned}
\label{eq:mpQP-linear-system-short-U}
\end{equation}
System~\eqref{eq:mpQP-linear-system-short-U} has infinitely many parametric solutions given by the affine expression
\begin{subequations}
\begin{equation} 
    x^*(p,y_2) = V_1\diag(\sigma_1)^{-1}U_1'(M_p p + M_1)+V_2y_2,\ y_2\in\rr^{n_x-n_M}
\label{eq:mpQP-solution-partial-rank}
\end{equation}
where $y_2$ is a free vector, if and only if 
\beqar 
    &&C_{-i}^{j_i}x_{-i}^*(p,y_2)+D_i^{j_i}p\leq e_i^{j_i},\ i=1,\ldots,N \label{eq:mpQP-CR-partial-rank}\\
    &&(U_2' M_p)p = -U_2' M_1.\label{eq:mpQP-CR-partial-rank-eq}
\eeqar
\end{subequations}
The critical region $CR_k$ is given by the projection
of the polyhedron defined by~\eqref{eq:mpQP-CR-partial-rank}--\eqref{eq:mpQP-CR-partial-rank-eq} onto the space of parameters $p$.
The conditions 
\begin{equation}
    U_2'M_p=0,\ U_2' M_1=0
\label{eq:U2}
\end{equation}
need to be satisfied for $CR_k$ to be full-dimensional and nonempty.
\begin{proposition}
Consider the linear system~\eqref{eq:mpQP-linear-system-short} with $\rank(M_x)=n_M<n_x$.
If conditions~\eqref{eq:U2} hold, then~\eqref{eq:mpQP-linear-system-short} is solvable
$\forall p\in\rr^{n_p}$ and, in particular, $\forall p\in CR_k$.
\end{proposition}
\begin{proof}
Note that, under~\eqref{eq:U2}, we have that
\[
    \begin{aligned}
    \rank([M_x\ |\ M_pp+M_1])&=\rank([U\diag(\sigma)V'\ |\ M_pp+M_1])\\
    &\hspace*{-2cm}=\rank([\diag(\sigma)V'\ |\ U'M_pp+U'M_1])\\
    &\hspace*{-2cm}=\rank\left[\begin{array}{ccc}
        \multicolumn{1}{c}{\begin{array}{c}\diag(\sigma_1)V_1'\end{array}} 
        & \vline & \begin{array}{c} U_1'M_{p}p+U_1'M_1 \end{array} \\
        \hline
        \multicolumn{1}{c}{\begin{array}{c}0\end{array}} 
        & \vline & \begin{array}{c} U_2'M_{p}p+U_2'M_1 \end{array}
        \end{array}\right]\\
    &\hspace*{-2cm}=n_M=\rank(M_x),\ \forall p\in\rr^{n_p}
    \end{aligned}
\]
which, by Rouche-Capelli theorem, implies that the linear system~\eqref{eq:mpQP-linear-system-short} is solvable
$\forall p\in\rr^{n_p}$.
\end{proof}


In the following, we present a variety of options for choosing a specific solution out of the infinite set.

\smallskip 

\subsubsection{Minimum norm solution}
A particular Nash equilibrium consists of choosing the equilibrium $x^*(p)$ that has minimum
Euclidean norm. This can be obtained by choosing $y_2$ as a function of $p$ as follows.
Since $V^\top$ is an orthogonal matrix, we have that
\[
    \begin{aligned}
    \|x^*(p,&y_2)\|_2^2 = \|V^\top x^*(p,y_2)\|_2^2 \\
    =& \|[V_1\ V_2]^\top(V_1\diag(\sigma_1)^{-1}U_1'(M_p p + M_1)+V_2y_2)\|_2^2\\
    =& \left\|\left[\begin{smallmatrix}
        V_1^\top V_1\diag(\sigma_1)^{-1}U_1'(M_p p + M_1)+V_1'V_2y_2\\
        V_2^\top V_1\diag(\sigma_1)^{-1}U_1'(M_p p + M_1)+V_2'V_2y_2
    \end{smallmatrix}\right]\right\|_2^2\\
    =& \left\|\left[\begin{smallmatrix}
        \diag(\sigma_1)^{-1}U_1'(M_p p + M_1)\\
        y_2
    \end{smallmatrix}\right]\right\|_2^2\\
    =& \|\diag(\sigma_1)^{-1}U_1'(M_p p + M_1)\|_2^2 + \|y_2\|_2^2.
    \end{aligned}
\]

Since the first term is independent of $y_2$, the minimum norm solution is obtained by minimizing $\|y_2\|_2^2$, yielding the mpQP problem:
\begin{equation}
    \begin{aligned}
    \change{y_2^*(p)} = \arg \min_{y_2} & \|y_2\|_2^2\\
    \st & C_{-i}^{j_i}x_{-i}^*(p,y_2)+D_i^{j_i}p\leq e_i^{j_i},\ i=1,\ldots,N
    \end{aligned}
\label{eq:mpQP-min-norm-solution}
\end{equation}
with parameter $p$ and decision variable $y_2\in\rr^{\change{n_x}-n_M}$. The corresponding critical regions $CR_{k,1},\ldots,CR_{k,m_k}\subseteq CR_k$
provide a subpartition of $CR_k$ into $m_k$ polyhedral regions, each one associated with a different affine solution
$\change{ y_2^*(p)}=C_{k,i}p+d_{k,i}$, $i=1,\ldots,m_k$. Finally, from~\eqref{eq:mpQP-solution-partial-rank}
we can retrieve the corresponding affine solution for $x^*(p)$ in each subregion $CR_{k,i}$ as follows:
\begin{equation} 
    x^*(p) = V_1\diag(\sigma_1)^{-1}U_1'(M_p p + M_1)+V_2(C_{k,i}p+d_{k,i}).
\label{eq:mpQP-solution-partial-rank-explicit}
\end{equation}



%

\smallskip 

\subsubsection{Welfare GNE solution} Another approach for selecting a GNE out of the infinitely many solutions is for a social planner to pick the one which maximizes a social welfare function or fairness criterion $f(x^*(p,y_2))$ as presented in~\cite{hall2025limits}. The resulting equilibrium $x^*$ would thus still be a point at which no agent has an incentive to deviate (strategically stable) while also optimizing some higher-level societal goals. Selecting the societally optimal GNE out of the set would result in the following multiparametric programming problem:
\begin{equation}
    \begin{aligned}
    \change{y_2^*(p)} = \arg \min_{y_2} &\,  f(x^*(p,y_2))\\
    \st & C_{-i}^{j_i}x_{-i}^*(p,y_2)+D_i^{j_i}p\leq e_i^{j_i},\ i=1,\ldots,N.
    \end{aligned}
\label{eq:mpQP-welfare-solution}
\end{equation}
The welfare metric $f$ must be such that problem~\eqref{eq:mpQP-welfare-solution} leads to PWA solutions over polyhedral critical regions; common cases are convex quadratic~\cite{bemporad2002explicit}, linear~\cite{BBM00mplp}, or functions that can be represented via mixed-integer linear programming~\cite{DP99b}. A common choice is the utilitarian sum $f^{SW}:\rr^{n_x}\to\rr$ defined as 
\begin{align}\label{eq:Util_sum}
 f^{SW}(x^*) = \sum_{i = 1}^{N} \frac{1}{2}(x^*)^\top Q_i x^*+(c_i+F_ip)^\top x^*.
\end{align}
Note that~\eqref{eq:mpQP-welfare-solution} selects the GNE which maximizes social welfare out of the set of GNEs; however, this does not imply it actually results in the social optimum. In fact, GNEs coincide with the social optimum when costs are the same across agents up to a constant, as discussed in~\cite{kulkarni2019efficiency}. Besides the utilitarian sum of costs, there exist a large variety of welfare metrics which are linear in the cost which have different interpretations, see~\cite{xinyingchen2023guide} for a detailed discussion.


\smallskip

\subsubsection{Variational GNE solutions}
\label{sec:variational-GNE}
For critical regions $CR_k$ characterized by infinitely many equilibria, we want to characterize a possible subregion 
$\overline {CR}_k\subseteq CR_k$ where {\it variational} GNE solutions exist. These are special solutions in two specific ways: ($i$) they correspond to the solution of a variational inequality and ($ii$) ensure that Lagrange multipliers associated with the coupling constraints are equal across agents~\cite{facchinei2009generalized, facchinei2009nash}, i.e., the sensitivities of the agents' costs to the coupling constraints are the same. 

In the explicit GNE framework, this corresponds to selecting $y_2$ as an affine function of $p$ over $CR_k$ to ensure consensus on the Lagrange multiplier \change{$\lambda_{c,i}$, i.e.\ the entry of the multiplier vector $\lambda_i$ in~\eqref{eq:KKT} associated with a coupling constraint $c$, }in the polyhedral region $CR_i^{j_i}$ of the $i$-th agent's best response in~\eqref{eq:mpQP-i}. The parametric multiplier function is piecewise-affine with respect to $(x_{-i},p)$~\cite[Theorem 2]{bemporad2002explicit}: 
\[
    \begin{aligned}
    \change{\bar \lambda_{c,i}}(x_{-i},p) = &\Lambda_x^{c,i,j}x_{-i} + \Lambda_p^{c,i,j}p + \Lambda_1^{c,i,j}
    \ \mbox{if}\ C_{-i}^jx_{-i}+D_i^jp\leq e_i^j\\
    &j=1,\ldots,N_i
    \end{aligned}
\]
\noindent  \change{where $\Lambda_x^{c,i,j}$, $\Lambda_p^{c,i,j}$ and $\Lambda_1^{c,i,j}$ are the affine coefficients of the multiplier $\bar \lambda_{c,i}$ on the $j$-th region, with possibly $\Lambda_x^{c,i,j}=\Lambda_p^{c,i,j}=\Lambda_1^{c,i,j}=0$ if $c$ is only weakly active in $CR_i^{j_i}$.}
Then, we can add the linear equations on top of~\eqref{eq:mpQP-linear-system}:
\begin{equation}
      \change{\bar\lambda_{c,i}}(x_{-i},p)=\change{\bar\lambda}_{c,i_1}(x_{-i_1},p),\ \forall i\in \mathcal{I}_c,i\neq i_1
\label{eq:lambda_i=lambda_i1}
\end{equation}
\noindent \change{where $\mathcal{I}_c$ is the set of agents involved in the coupling constraint $c$
and $i_1$ is the first element of $\mathcal{I}_c$,}
and solve the complete system with respect to $x$ and parameter $p$ (for numerical efficiency, we could also solve the linear system $\Lambda_x^{c,i,j}x^*_{-i}(p,y_2) + \Lambda_p^{c,i,j}p + \Lambda_1^{c,i,j}=\Lambda_x^{c,i_1,j}x^*_{-i_1}(p,y_2) + \Lambda_p^{c,i_1,j}p + \Lambda_1^{c,i_1,j}$, $\forall i\in \mathcal{I}_c,i\neq i_1$, for $y_2$ parametrically with respect to $p$.)

Three cases can occur: ($i$) the parametric linear system has a unique equilibrium solution 
with the associated critical region $\overline{CR}_k$ computed as in~\eqref{eq:mpQP-solution-full-rank-CR}; ($ii$) no solution exists, meaning that no v-GNE solution exists; ($iii$) infinitely many solutions, meaning that infinitely many v-GNE solutions exist over the critical region $\overline{CR}_k$ obtained by projecting over the $p$-space. In this case, one can proceed as described in the previous sections to define a minimum-norm or welfare GNE solution, if desired. It is immediate to prove that,
due to the added constraint~\eqref{eq:lambda_i=lambda_i1}, $\overline{CR}_k\subseteq CR_k$.

A possible way of partitioning of $CR_k\setminus\overline{CR}_k$ into polyhedral subregions is to solve the following mpQP problem with decision variable $y_2\in\rr^{\change{n_x}-n_M}$ and parameter $p$ over $CR_k$:
\begin{equation}
\small
    \begin{aligned} 
  \min_{y_2} & \sum_{c}\sum_{i\in \mathcal{I}_c,i\neq i_1} \|\change{\bar\lambda_{c,i}}(x_{-i}^*(p,y_2),p)-\change{\bar\lambda}_{c,i_1}(x_{-i_1}^*(p,y_2),p)\|_2^2\\
        \st & C_{-i}^{j_i}x_{-i}^*(p,y_2)+D_i^{j_i}p\leq e_i^{j_i},\ i=1,\ldots,N
    \end{aligned}
\label{eq:mpQP2-vgne}
\end{equation}
The critical region of problem~\eqref{eq:mpQP2-vgne} corresponding to the null combination of active constraints provides the subregion $\overline{CR}_k$, where the optimal cost of~\eqref{eq:mpQP2-vgne} is zero as~\eqref{eq:lambda_i=lambda_i1} can be satisfied within $\overline{CR}_k$. Note that, instead, the other critical regions of~\eqref{eq:mpQP2-vgne}, in general, do not provide v-GNE solutions. Therefore, other v-GNE solutions may exist in $CR_k\setminus \overline{CR}_k$, that we do not explicitly characterize.

\subsection{Example 1 revisited}

\begin{figure}
    \centering
\begin{subfigure}{0.8\columnwidth}
    \centering
    \includegraphics[width = 0.8\linewidth]{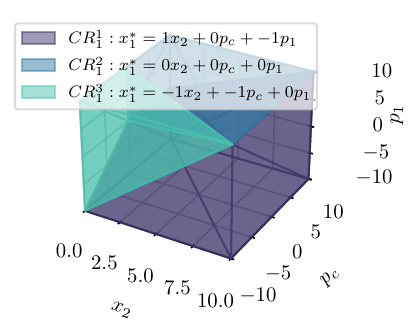}
\end{subfigure}
\begin{subfigure}{0.8\columnwidth}
    \centering
    \includegraphics[width = 0.8\linewidth]{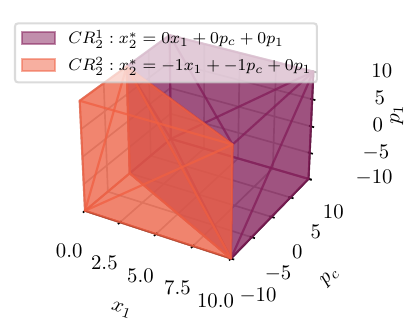}
\end{subfigure}
\caption{Critical regions of each agent's mpQP of the running example.}\label{fig:CriticalRegionsPerAgent}
\end{figure}    

\begin{figure}
    \centering
    \begin{subfigure}{0.23\textwidth}
        \centering
        \includegraphics[width=\linewidth]{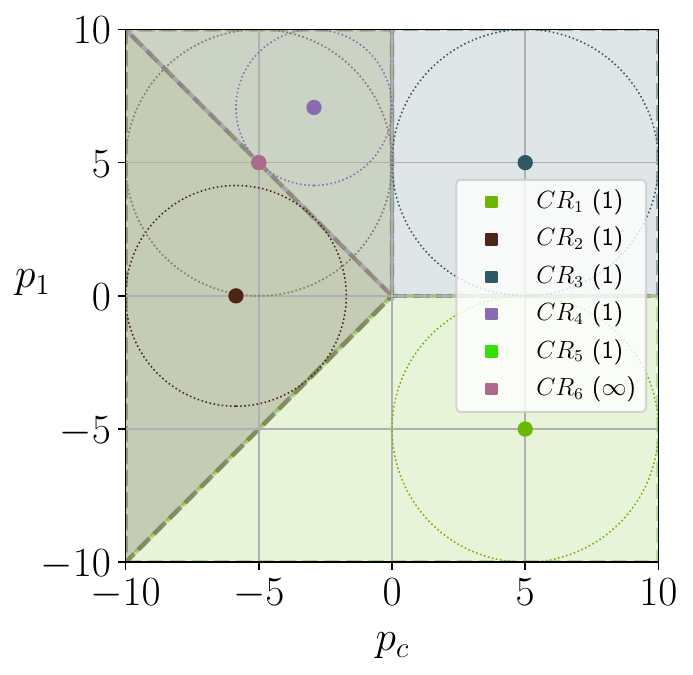}
        \caption{Critical regions (no split).}\label{fig:CRfull}
    \end{subfigure}\\
        \begin{subfigure}{0.22\textwidth}
        \centering
        \includegraphics[width=\linewidth]{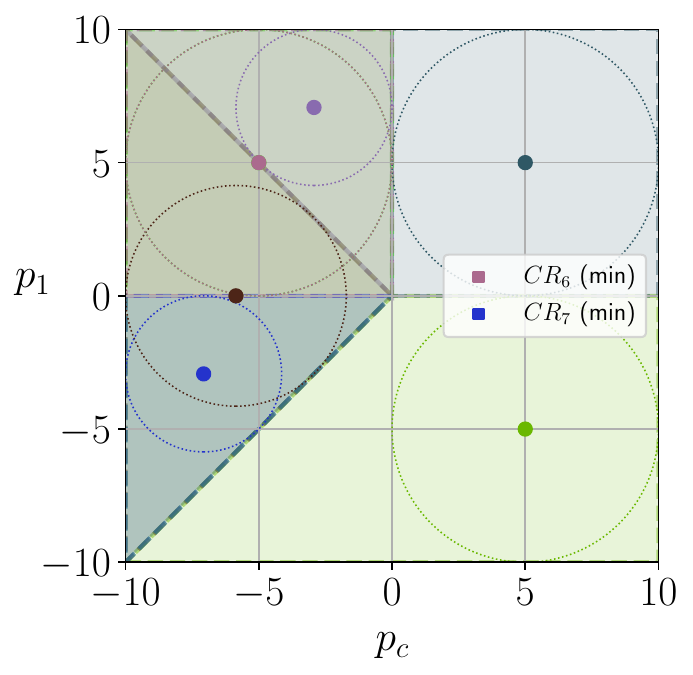}
        \caption{Minimum-norm solution~\eqref{eq:mpQP-min-norm-solution}}\label{fig:CRminNorm}
    \end{subfigure}
    \begin{subfigure}{0.22\textwidth}
        \centering
        \includegraphics[width=\linewidth]{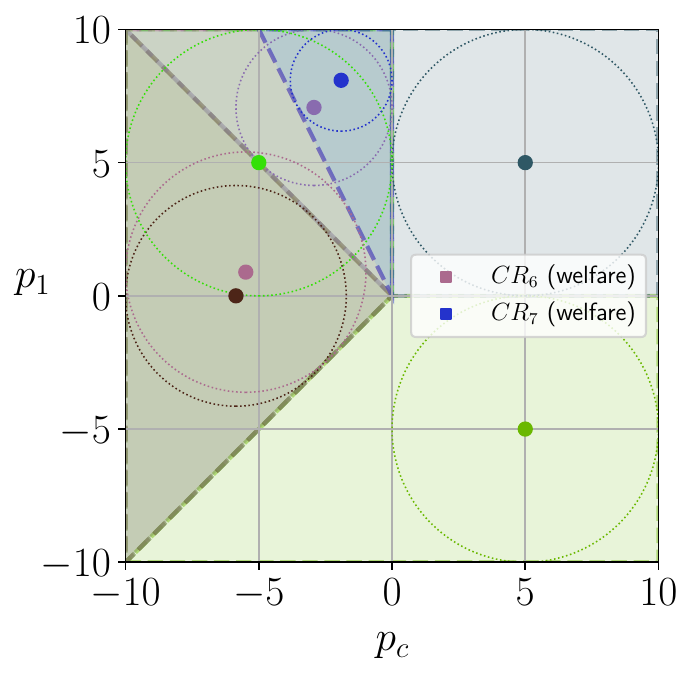}
        \caption{Welfare solution~\eqref{eq:mpQP-welfare-solution}}\label{fig:CRwelfare}
    \end{subfigure}
    \caption{The critical regions of the explicit GNE solution for the running example.}\label{fig:NECriticalRegions}
\end{figure}

\change{To demonstrate the mechanism of the multiparametric GNE algorithm we now present a step-by-step solution to the Example~1 posed in Section~\ref{sec:MultiparametricGNEs}. We start by writing out \eqref{eq:mpQP-i} and thus identify the per-agent critical regions $CR_i^j$, which are also shown in Figure~\ref{fig:CriticalRegionsPerAgent}:
\[
\begin{array}{llr}
\bar x_1 =x_2 -p_1: & CR_1^1=\{0 \leq x_2,\ p_1 \leq x_2,~ \frac{(p_1 - p_c)}{2}\leq x_2 \}\\
\bar x_1 = 0:  &CR_1^2 = \{0 \leq x_2 \leq p_1,~  -x_2\leq p_c \}\\
\bar x_1 = -x_2-p_c: & CR_1^3 =  \{0\leq x_2\leq \frac{(p_1 - p_c)}{2},~ x_2\leq-p_c \}\\[3ex]
\bar x_2 = 0: & CR_2^1 = \{ 0 \leq x_1,\ -p_c \leq x_1 \}\\
\bar x_2 = -x_1 -p_c:  &CR_2^2 =  \{  0 \leq x_1 \leq -p_c \}. \\
&&
\end{array}
\]
Next, to derive~\eqref{eq:mpQP-solution}, we combine the critical regions of each agents' mpQP 
and eliminate empty regions. This leads to the following critical regions of the parametric GNEP, depicted in Figure~\ref{fig:CRfull}:

%

\begin{itemize}[leftmargin=*]
\item $CR_1$ ($CR_1^1$ = unconstrained, $CR_2^1$ = local constraint $A_3x=b_3+S_3p$ active):
$p_1\leq x_2\geq 0,\ \frac{(p_1 - p_c)}{2}\leq x_2,\ 0\leq  x_1\geq  -p_c$,
leading to $CR_1=\{p:\ p_1\leq 0,\ p_1\leq p_c\}$;
\item $CR_2$ ($CR_1^2$ = local constraint $A_2x=b_2+S_2p$ active,
$CR_2^1$ = local constraint $A_3x=b_3+S_3p$ active, $x_1 =x_2 = 0$): 
$0\leq x_2 \leq p_1,  -x_2\leq p_c  ~;~ 0\leq  x_1, -p_c\leq x_1$,
leading to $CR_2=\{p:\ 0\leq p_1,\ 0\leq p_c\}$;
\item $CR_3$ ($CR_1^3$ = coupling constraint $A_1x=b_1+S_1p$ active,
$CR_2^1$ = local constraint $A_3x=b_3+S_3p$ active, $x_2 = 0$): 
$x_2\leq \frac{(p_1 - p_c)}{2},\ x_2\leq -p_c\ ;\ 0\leq x_1,\ -p_c\leq x_1$,
leading to $CR_3=\{p:\ p_c\leq 0,\ p_c\leq p_1\}$;
\item $CR_4$ ($CR_1^1$ = unconstrained,
$CR_2^2$ = coupling constraint $A_1x=b_1+S_1p$ active):
$p_1\leq x_2,\ \frac{(p_1 - p_c)}{2}\leq x_2\ ;\ x_1\leq -2p_c,\ x_1\leq -p_c$,
leading to $CR_4=\{p:\ p_c\leq -p_1,\ p_c\leq p_1\}$;
\item $CR_5$ ($CR_1^2$ = local constraint $A_2x=b_2+S_2p$ active, $x_1 = 0$,
$CR_2^2$ = coupling constraint $A_1x=b_1+S_1p$ active): 
$x_2\leq p_1,\ -x_2\leq p_c\ ;\ x_1\leq -2p_c,\ x_1\leq -p_c$,
leading to $CR_5=\{p:\ p_c\leq 0,\ -p_c\leq p_1\}$;
\item $CR_6$ ($CR_1^3$ and $CR_2^2$ both corresponding to coupling constraint $A_1x=b_1+S_1p$
active): 
$0\leq x_2\leq \frac{(p_1 - p_c)}{2},\ x_2\leq-p_c,\ 0 \leq x_1 \leq -p_c$,
leading to $CR_6=\{p:\ p_c\leq 0,\ p_c\leq p_1\}$.
\end{itemize}}

\begin{table}
\centering
\setlength{\tabcolsep}{3pt}
\caption{\change{Critical regions with unique solutions for the example.}}
\label{tab:unique_critical_regions}\textcolor{black}{
\begin{tabular*}{0.85\columnwidth}{@{\extracolsep{\fill}}cccc@{}}
\toprule
$CR_k$ & $M_x$ & $M_p$ & $x^*$ \\
\midrule
$CR_1$ & $\left[\begin{smallmatrix}1&-1\\0&1\end{smallmatrix}\right]$ & $\left[\begin{smallmatrix}0&-1\\0&0\end{smallmatrix}\right]$ & $\left[\begin{smallmatrix}-p_1\\0\end{smallmatrix}\right]$ \\[4pt]
$CR_2$ & $I_2$ & $\left[\begin{smallmatrix}0&0\\0&0\end{smallmatrix}\right]$ & $\left[\begin{smallmatrix}0\\0\end{smallmatrix}\right]$ \\[4pt]
$CR_3$ & $\left[\begin{smallmatrix}1&1\\0&1\end{smallmatrix}\right]$ & $\left[\begin{smallmatrix}-1&0\\0&0\end{smallmatrix}\right]$ & $\left[\begin{smallmatrix}-p_c\\0\end{smallmatrix}\right]$ \\[4pt]
$CR_4$ & $\left[\begin{smallmatrix}1&-1\\1&1\end{smallmatrix}\right]$ & $\left[\begin{smallmatrix}0&-1\\-1&0\end{smallmatrix}\right]$ & $\tfrac{1}{2}\left[\begin{smallmatrix}-p_c-p_1\\p_1-p_c\end{smallmatrix}\right]$ \\[4pt]
$CR_5$ & $\left[\begin{smallmatrix}1&0\\1&1\end{smallmatrix}\right]$ & $\left[\begin{smallmatrix}0&0\\-1&0\end{smallmatrix}\right]$ & $\left[\begin{smallmatrix}0\\-p_c\end{smallmatrix}\right]$ \\
\bottomrule
\end{tabular*}}
\end{table}

\change{
We summarize the parametric GNE for all regions with unique solutions in Table~\ref{tab:unique_critical_regions}. In critical region $CR_6$, we have infinitely many solutions, as
the linear system~\eqref{eq:mpQP-linear-system-short} is given by
$\left[\begin{smallmatrix}
1 & 1\\
1  & 1
\end{smallmatrix}\right] x = \left[\begin{smallmatrix}
-1  &0 \\
-1 & 0  
\end{smallmatrix}\right] p 
$ and, therefore, $\rank(M_x) = 1<n_x=2$. 
We want to derive the explicit expression of $x^*$ as a function of $y_2$ and rounded to the third decimal for simplicity of exposition. The SVD of $M_x$ gives: 
\begin{align*}
U ~= \left[\begin{smallmatrix} -0.707 &  -0.707\\
       -0.707 &  0.707 \end{smallmatrix}\right],\ \sigma = \left[\begin{smallmatrix} 2 \\ 0
        \end{smallmatrix}\right],\ V = 
        \left[\begin{smallmatrix}-0.707 &  -0.707\\
       -0.707 &  0.707 \end{smallmatrix}\right].
\end{align*}
Following \eqref{eq:mpQP-linear-system-short-U}, we take $y = V^\top x$ and left-multiply by $U^\top$
to get
$
2 y_1 = \left[\begin{smallmatrix} -0.707 \\  -0.707 \end{smallmatrix}\right] ^\top  \left[\begin{smallmatrix}
-1  &0 \\
-1 & 0  
\end{smallmatrix}\right] p$,
$0 = \left[\begin{smallmatrix} -0.707 \\  0.707 \end{smallmatrix}\right] ^\top  \left[\begin{smallmatrix}
-1  &0 \\
-1 & 0  
\end{smallmatrix}\right] p
$, 
which gives infinitely many solutions parametrized in $y_2$:
\begin{equation}
\begin{aligned}
x^*(p,y_2) &= V_1 \diag(\sigma_1)^{-1} U_1^\top (M_p p + M_1) + V_2 y_2\\
 &= \left[\begin{smallmatrix} -0.5 \\ -0.5 \end{smallmatrix}\right] p_c  +\left[\begin{smallmatrix} -0.707 \\ 0.707\end{smallmatrix}\right]  y_2.
\end{aligned}
\label{eq:mpQP-infinitely-many-solutions}
\end{equation}
Figure~\ref{fig:CRfull} shows the GNE critical regions without further splitting, Figure~\ref{fig:CRminNorm} under the minimum-norm solution, and Figure~\ref{fig:CRwelfare} under the welfare solution. Table~\ref{tab:region_split} summarizes the parametric GNE laws for the split regions.}

\begin{table}
\centering
\setlength{\tabcolsep}{6pt}
\caption{\change{GNE solutions on the subpartition of $CR_6$.}}
\label{tab:region_split}\textcolor{black}{
\begin{tabular*}{0.8\columnwidth}{@{\extracolsep{\fill}}ccc@{}}
\toprule
$CR_k$ & $x^*$ min-norm~\eqref{eq:mpQP-min-norm-solution} & $x^*$ welfare~\eqref{eq:mpQP-welfare-solution} \\
\midrule
$CR_6$ & $\left[\begin{smallmatrix}-\frac{p_c}{2}\\ -\frac{p_c}{2}\end{smallmatrix}\right]$ & $\frac{1}{3}\left[\begin{smallmatrix}-2 & -1\\ -1 & 1\end{smallmatrix}\right]p$ \\[6pt]
$CR_7$ & $\frac{1}{2}\left[\begin{smallmatrix}-1 & -1\\ -1 & 1\end{smallmatrix}\right]p$ & $\left[\begin{smallmatrix}0\\ -p_c\end{smallmatrix}\right]$ \\
\bottomrule
\end{tabular*}}
\end{table}

%

\subsection{Multiparametric GNE Algorithm}
In light of the developments of the previous sections, we summarize all the steps required for computing an explicit GNE solution to a given quadratic GNEP as in~\eqref{eq:mpQPi} in Algorithm~\ref{alg:ParametricGNE}.
Note that the polyhedron in $(p,y_2)$-space given by~\eqref{eq:mpQP-CR-partial-rank} is stored at Line~\ref{step:p-y2} to determine, for a given parameter $p$, the set of admissible values $y_2$ defining the infinitely many GNE solutions in $CR_k$. A Python implementation of the algorithms presented in this paper is available on \url{https://github.com/bemporad/nash_mpqp}.

\begin{algorithm}
\caption{Multiparametric Generalized Nash Equilibrium}
\label{alg:ParametricGNE}
\begin{algorithmic}[1]
\State \textbf{Input:} GNEP with $N$ agents \eqref{eq:mpQPi}, parameters $p \in \PP \subseteq \rr^{n_p}$.
\State \textbf{Output:} PWA (possibly non-unique) GNE solution $x^*$.
\Statex
\State \textbf{Step 1}: {\it Solve mpQP \eqref{eq:mpQP-i} for each agent}
\For{$i = 1, \ldots, N$}
     \State Solve: \change{$\bar x_i$}$(x_{-i},p) = \arg\min_{x_i} \frac{1}{2}x^\top Q_i x + (c_i + F_i p)^\top x$
    \Statex \hspace{\algorithmicindent} subject to $Ax \leq b + Sp$; 
    \State Get explicit laws \change{$\bar x_i$}$(x_{-i},p)$ and regions $CR_i^j$ as in~\eqref{eq:mpQP-i};
\EndFor

\Statex
\State \textbf{Step 2}: {\it Combine explicit laws \& regions \eqref{eq:mpQP-solution}}
\For{each combination $\CC_k$ of critical region indices}
    \If{$CR_k \neq \emptyset$}
        \State Store $CR_k$ and parametric linear system $(M_x^k, M_p^k, M_1^k)$ 
        \Statex \hspace{\algorithmicindent}\hspace{\algorithmicindent} as in~\eqref{eq:mpQP-linear-system-short} (and, possibly~\eqref{eq:lambda_i=lambda_i1});
        \If{$\rank(M_x) < n_x$} \Comment{Infinitely many solutions}
            \State Store polyhedron in $(p,y_2)$-space given by~\eqref{eq:mpQP-CR-partial-rank};\label{step:p-y2}
            \State If split, solve mpQP~\eqref{eq:mpQP-min-norm-solution} (min norm),~\eqref{eq:mpQP-welfare-solution} (welfare),
            \Statex \hspace{\algorithmicindent}\hspace{\algorithmicindent}\hspace{\algorithmicindent} or~\eqref{eq:mpQP2-vgne} (v-GNE) over $CR_k$;
        \EndIf
    \EndIf
\EndFor

\State \Return $x^*(p)$ (or $x^*(p,y_2)$) for all (sub)regions found.
\end{algorithmic}
\end{algorithm}

\section{Explicit game-theoretic MPC} \label{sec:GameTheoreticMPC}

A direct application of the explicit GNE algorithm presented in Section~\ref{sec:MultiparametricGNEs} is that of explicit game-theoretic MPC. In line with the constrained linear-quadratic game in~\eqref{eq:mpQPi}, agents minimize a quadratic cost over a finite horizon $K$ subject to linear time-invariant dynamics and polyhedral constraints as follows:

\begin{subequations}\label{explicitMPC}
\begin{align}
\min_{u_{i}, x} &\sum_{k=0}^{K-1} x_{k}^\top \change{\mathcal Q}_i x_{k}+u_{k}^\top \change{\mathcal R}_i u_{k} + ({\mathcal F}_i^x p)^\top x_k + ({\mathcal F}_i^u p)^\top u_k\label{explicitMPC_cost}\\
\st &x_{k+1} = \change{\mathcal A} x_k +\sum_{j=1}^{N}\change{\mathcal B}_{j} u_{j,k} + W_d p\label{explicitMPC_dynamics}\\ 
&C^u_{i,k}  u_{i,k}\leq c^u_i + {\mathcal S}^u_i p \label{explicitMPC_inputConstr}\\[1ex]
& C^u  u_k \leq c^u + {\mathcal S}^u p \label{explicitMPC_inputcouplConstr}\\[1ex]
&   C^x x_k\leq c^x + {\mathcal S}^x p \label{explicitMPC_statecouplConstr} \\[1ex]
& x_0 = W_x p. \label{explicitMPC_initialCond}
\end{align}
\end{subequations}
\change{where all stage constraints hold for $k\in  \{0,1,\ldots,K-1\}$}, $\change{\mathcal A}$ is the global state transition matrix, $\change{\mathcal B}_j$ the input matrix of agent $j$,  \change{and $W_d$, $W_x$ map the parameter $p$ to the disturbance and initial condition. The matrices $C^u_{i,k}, c^u_i, {\mathcal S}^u_i$ define agent $i$'s local input constraints, $C^u, c^u, {\mathcal S}^u$ and $C^x, c^x, {\mathcal S}^x$ the coupling input and state constraints, and $F_i^x, F_i^u$ the parameter dependent linear cost terms, all in line with~\eqref{eq:mpQPi}.} The cost function of each agent is quadratic in the global state $x_k$ and a function of all agent inputs $u_k$ weighted with $\change{\mathcal Q}_i$ and $\change{\mathcal R}_i$, respectively. Note that the problem formulation in~\eqref{explicitMPC} allows for the GNE problem to be parametrized in the initial condition $x_0$, the linear cost, reference signals, disturbances acting on the linear dynamics, as well as the right hand side of the local and coupling constraints. In particular, set-point tracking problems can be formulated by including linear terms in the cost function as in~\eqref{explicitMPC_cost},
as minimizing $(x_k-x_r)^\top \change{\mathcal Q}_i(x_k-x_r)+(u-u_r)^\top \change{\mathcal R}_i(u_k-u_r)$ is equivalent to minimizing $x_k^\top \change{\mathcal Q}_i x_k + u_k^\top \change{\mathcal R}_i u_k - 2 x_r^\top \change{\mathcal Q}_i x_k - 2 u_r^\top \change{\mathcal R}_i u_k$, with state and input setpoints $x_r, u_r$ as part of the parameters $p$.

\subsection{Battery Charging Game}
In the following we consider a game-theoretic MPC example, specifically a battery charging game which, similar to the one presented in~\cite{hall2024stability}, is given by the following GNE problem:
\begin{subequations}
\label{eq:MPCPerAgent}
\begin{align}
\label{eq:RunningCost}
\displaystyle \min_{u_i,\, x_i}  &\;\sum_{k= 0}^{K-1} \ell_i(x_{i,k}, u_{k})\\
\textrm{s.t.} \quad &  x_{i, k+1} =  \change{\mathcal A}_i x_{i,k} + \change{\mathcal B}_i u_{i,k} \label{eq:Constr1}\\
&l_{i,k} = u_{i,k}+ d_{i,k} \label{eq:Constr2}\\
& 0			\leq  l_{i,k} 	\leq 		l_{i,\text{max}}\label{eq:Constr3}\\
& -u_{i,\text{max}}  \leq  u_{i,k}  \leq u_{i,\text{max}}  \label{eq:Constr4}\\ 
&0 \leq  \textstyle \sum_{j= 1}^N l_{j,k} \leq L_{\text{max}} \label{eq:Constr5}\\
&\; x_{i,0} = \mathbf{x}_i. \label{eq:Constr6}
\end{align}
\end{subequations}
\change{where the constraints hold for $k\in  \{0,1,\ldots,K-1\}$},  and we consider linear charging dynamics in~\eqref{eq:Constr1} \change{with measured initial state $\mathbf{x}_i$} in~\eqref{eq:Constr6}, and leakage rate and charging efficiency $\change{\mathcal A}_i, \change{\mathcal B}_i \!\in\! \left[0, 1\right]$.  The battery charging/discharging input $u_{i,k}$ is constrained by the charging rate $ u_{i,\text{max}}\geq 0$. Further, the power a user buys from the grid $l_{i,k}$ is a function of their storage $u_{i,k}$ and demand $d_{i,k}$ as defined in~\eqref{eq:Constr2} and has an upper limit $l_{i, \text{max}}$. The aggregate load of all consumers connected to the same point of common coupling is constrained by $L_{\text{max}}>0$ in~\eqref{eq:Constr5}.

Each self-interested consumer $i = 1\dots N$ aims to minimize its electricity bill as well as the operational cost of its battery, yielding a local cost function of the form
\begin{align*} 
\ell_i(x_{i,k},u_{k}) &=\,  \underbrace{\left(\gamma^{1}_i \sum_{j = 1}^{N}l_{j,k}+ \gamma^{2}_{i,k} \right) \,l_{i,k}}_{\text{energy cost}} +\underbrace{\gamma^3_i  \, (x_{i,k}-x^{\text{ref}}_i)^2}_{\text{battery usage}},
\end{align*}
where $l_{i,k} = u_{i,k}+ d_{i,k}$ is the power bought from the grid~\eqref{eq:Constr2}. We define the parameters of the multiparametric GNEP as \change{$p = [\mathbf{x}_1,\ldots,\mathbf{x}_N,\ \gamma_1^2,\ldots,\gamma_N^2,\ L_{\text{max}}]^\top \in \mathbb{R}^{2N+1}$}.  Note that this parametrization is inspired by a real application in energy management, developed jointly with the local DSO of the canton Lucerne, Switzerland. For energy management applications it is important to reduce computational requirements in each household as this requires more hardware and increases the number of potential failure points, and thus maintenance overhead. In addition, using an explicit solution minimizes real-time communication requirements, as distributed GNE seeking methods require many iterations and thus communication rounds  between households sharing entire input, state and dual variable trajectories, which leads to privacy concerns and is fragile. With the proposed approach, households only need to share the initial condition $\mathbf{x}_i$ in real-time. The DSO can reduce aggregate consumption by sending a new load limit $L_{\text{max}}$, and the base energy price $\gamma_i^2$ of each household can be updated, e.g., for time-of-use pricing as implemented in the canton of Lucerne from 2025.

\smallskip 

\begin{table}
\setlength{\tabcolsep}{5.1pt} 
\begin{tabular}{@{}lllllllll@{}}
\toprule
 & $\gamma_i^1$ & $\gamma_i^3$ & $u_{i,\max}$ & $l_{i,\max}$ & $A_i$ & $B_i$ & $d_i$ & $x_i^{\text{ref}}$ \\ \midrule
Agent 1 & 0.1 &0.05 & 10 & 10  &0.960 & 0.71 & 1.67 & 15  \\
Agent 2 & 0.01 &0.01&  10 & 10  & 0.985 &0.76 & 1.27 & 15  \\
\bottomrule
\end{tabular}
\caption{Parameters for the Battery charging GNE problem~\eqref{eq:MPCPerAgent}}\label{tab:gne_parameters}
\end{table}

\subsubsection{Simulation results}
\begin{figure}
    \centering
    \begin{subfigure}{\columnwidth}
        \centering
    \includegraphics[width = \linewidth]{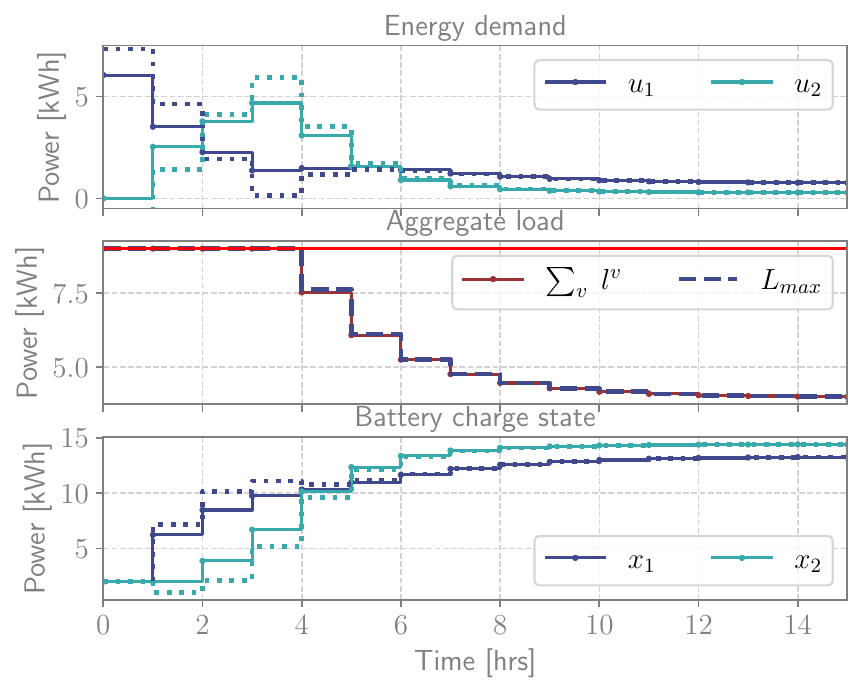}
    \caption{Closed-loop simulation comparing the explicit welfare solution (dotted lines) to the v-GNE solution (solid lines).} \label{fig:Battery-GNE-MPC-trajectories}
\end{subfigure}    
    \begin{subfigure}{\columnwidth}
        \centering
    \includegraphics[width = 0.7\linewidth]{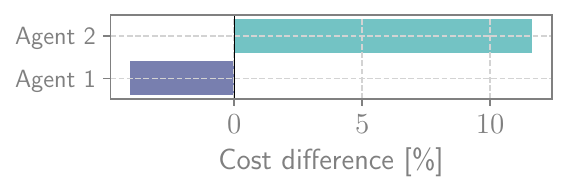}
      \caption{The cost increase for the explicit welfare solution compared to the v-GNE solution over a horizon of 14 time steps.} \label{fig:Battery-GNE-MPC-cost}
\end{subfigure}
\caption{Closed-loop simulation comparing the explicit welfare solution (dotted lines) to a v-GNE solution (solid lines) solved with a centralized solver.}\label{fig:Battery-GNE-MPC}
\end{figure}

We simulate the closed-loop system under the explicit feedback law for the welfare solution and compare it to the v-GNE solution for a horizon of $K=\change{4}$, $\gamma^2_1 = \gamma_2^2=0.01$ and $L_{\text{max}} = 9$. Additional parameter values are reported in Table~\ref{tab:gne_parameters}. The closed-loop trajectories are given in Figure~\ref{fig:Battery-GNE-MPC-trajectories}; as expected, charging patterns differ between the welfare and variational solution which influences the steady-state both agents converge to. Figure~\ref{fig:Battery-GNE-MPC-cost} shows the closed-loop cost difference in the welfare GNE setting compared to the variational setting. Specifically, we demonstrate that Agent 2 pays over 15\% more in the welfare GNE closed-loop, which only decreases Agent 1's cost by less than 1\% but yields a lower total cost. This is a known issue: minimizing the utilitarian sum~\eqref{eq:Util_sum} can disproportionately affect individual agents as their self-interest is not protected. Yet, in contrast, the v-GNE solution is generally not Pareto optimal and can arbitrarily deteriorate social welfare~\cite{kulkarni2019efficiency}. This demonstrates two important advantages of our proposed explicit multiparametric GNE method: ($i$) ({\it A priori selection}) a performance metric or welfare measure can be defined a priori to select the GNE, overcoming the limitation of existing algorithms to only select and converge to the v-GNE solution; ($ii$) ({\it Interpretability}) the outcomes for individual agents can be scrutinized before deployment for all possible parameter values guiding the decision which of the GNE solutions to select. This is especially relevant in closed-loop systems which run for infinite time and affect users on a daily basis, such as in energy management~\cite{hall2025limits}.

\change{Table~\ref{tab:gnep_stats} gives the increase of the number of per-agent regions and critical regions of the GNEP as a function of the horizon length $K$. As defined previously, $N_i$ denotes the per-agent critical region count, $\prod_{i=1}^N N_i$ the number of best-response combinations $\CC_k$, and $\#\{CR_k\}$ the cardinality of the final partition obtained after splitting critical regions of infinitely many equilibria into their subregions $CR_{k,1},\ldots,CR_{k,m_k}$ and removing empty combinations. Note that for $K=2$  $\prod_i N_i<\#\{CR_k\}$ due to the additional splitting regions whereas for $K\geq3$ many combinations are empty and get removed and thus $\prod_i N_i>\#\{CR_k\}$.}

\begin{table}
\centering 
\small
\caption{\textcolor{black}{Explicit GNEP solve statistics for the battery example for two agents $N=2$ with increasing prediction horizon $K$.}}
\label{tab:gnep_stats}\textcolor{black}{
\begin{tabular}{r r r r r}
\toprule
$K$ &  $t_{\mathrm{offline}}$ [s] & $N_i$ & $\prod_i N_i$ & $\#\{CR_k\}$\\
\midrule
2 & 4.52    & 9, 12    & 108   & 132  \\
3 & 37.65   & 30, 31   & 930   & 706   \\
4 & 261.17  & 74, 74   & 5476  & 3113 \\
5 & 1576.09 & 179, 162 & 28998 & 8903  \\
\bottomrule
\end{tabular}}
\end{table}

\subsection{Two-mass-spring-damper system}
\begin{figure}
    \centering
    \includegraphics[width=0.57\linewidth]{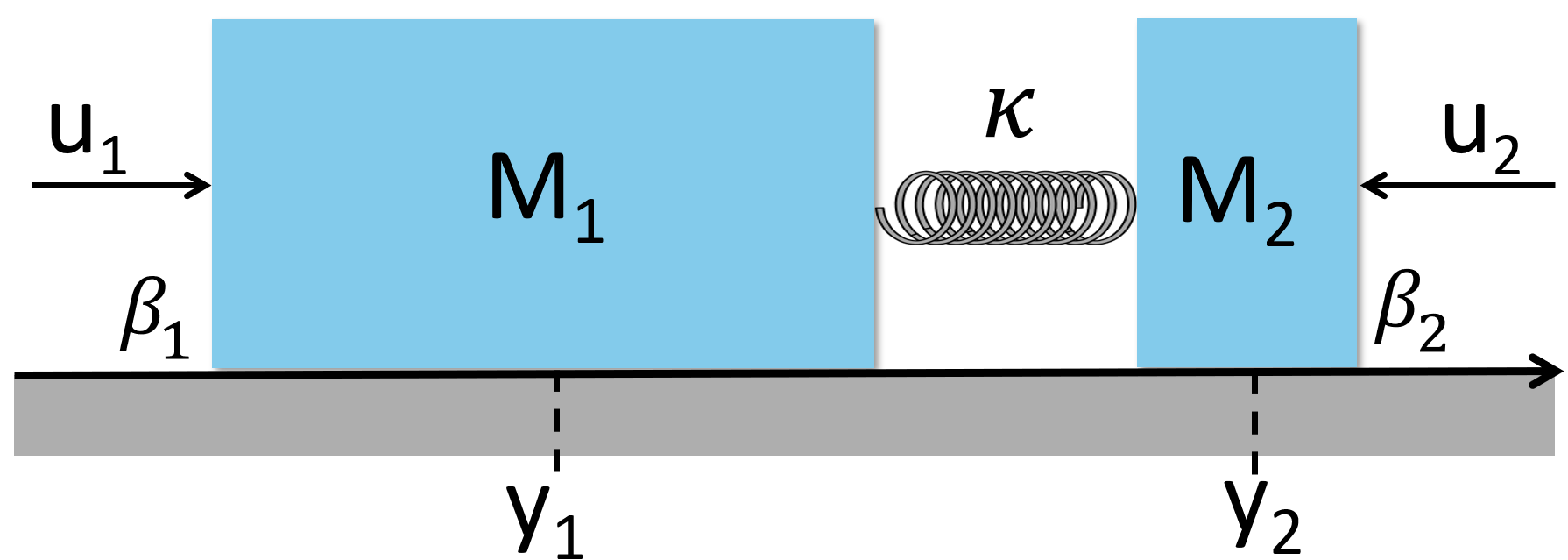}
    \caption{Two-mass-spring-damper system: each agent applies force $u_i$ to drive its mass to a reference $r_i$, under heterogeneous MPC costs and common position constraints.}
    \label{fig:two-mass-spring-damper}
\end{figure}

Consider the simple mechanical system consisting of two masses $M_1$, $M_2$ connected by a spring of constant $\kappa$ shown in Figure~\ref{fig:two-mass-spring-damper}. Each mass is subject to viscous friction, with coefficient $\beta_i$ and is controlled by an agent applying a force input $u_i$ to move the mass position $y_i$ to a desired position $r_i$, $i=1,2$. Each input $u_i$ is generated by its own linear MPC controller, each one based on a different cost function, that share the same model and position constraint $y_2\geq y_1+\Delta y$. The prediction model is obtained by discretizing the continuous-time dynamics 
\[
    \begin{aligned}
    M_1 \ddot{y}_1 &= -\kappa(y_1-y_2) - \beta_1 \dot{y}_1 + u_1\\
    M_2 \ddot{y}_2 &= \kappa(y_1-y_2) - \beta_2 \dot{y}_2 - u_2
    \end{aligned}
\]
with sampling time $T_s$ and zero-order-hold on the inputs. The resulting model is used by each agent to  minimize the following costs over a prediction horizon of $K$ steps:
\[
    \begin{aligned}
    J_1 = &\sum_{k=0}^{K-1} \|\change{\xi}_{1,k}-\change{\xi}_{2,k}\|_{\change{\mathcal{Q}}_1}^2 +\|\Delta u_{1,k}+\Delta u_{2,k}\|_{R_1}^2\\
    J_2 = &\sum_{k=0}^{K-1} \|\change{\xi}_{2,k}\|_{\change{\mathcal{Q}}_2}^2 +\|\Delta u_{2,k}\|_{R_2}^2
    \end{aligned}
\]
where $\change{\xi}_{i,k} = y_{i,k}-r_i$ are the tracking errors, $\Delta u_{i,k}=u_{i,k}-u_{i,k-1}$ the input increments, and $\|x\|_\change{\mathcal{Q}}^2 = x^\top \change{\mathcal{Q}} x$. The first agent aims to smoothly change the  applied total force $u_1+u_2$ and have equal tracking errors, while the second agent aims to track its own reference position with smooth changes of its own actuator. The parameters of the system are $M_1 = 3$ kg, $M_2 = 1$ kg, $\kappa = 0.5$, $\beta_1 = 1.5$, $\beta_2 = 1.0$, $T_s = 0.2$, and $\Delta y = 0.5$ (MKS units), the MPC parameters are $Q_1 = \smallmat{1\ -1\\-1\ 1}$, $R_1=\smallmat{1\ 1\\1\ 1}$, $Q_2 = \smallmat{0\ 0\\0\ 1}$, $R_2 = \smallmat{0\ 0\\0\ .1}$, $K=10$. Position constraints are enforced over a shorter horizon of 3 prediction steps.
For comparison, we also consider an \change{optimization-based MPC controller which minimizes a single joint objective, the sum of costs $J_1+J_2$, by controlling both inputs simultaneously subject to the same constraints.}

We run Algorithm~\ref{alg:ParametricGNE} to compute the explicit GNE solution with parameter $p = [x(t)^\top\ u(t-1)^\top\ r_1(t)\ r_2(t)]^\top\in\rr^8$ over the range $\|p\|_\infty\leq 100$, where $x(t)$ is the current state of the system at time $t$ and $u(t-1)$ the previous input applied. The explicit solution consists \change{of 15 critical regions, 12 with unique solutions and 3 with infinitely many solutions. By applying the minimum-norm criterion in~\eqref{eq:mpQP-min-norm-solution}, we obtain a unique explicit GNE solution with 27 critical regions}.  \change{The offline computation time was 12.28~s
on an Apple M4 Max processor with 64~GB of RAM. The online evaluation time of the explicit GNE solution ranges between 0.008 and 0.069~ms, with an average of 0.012~ms. For comparison, we solve the GNE problem online via the MILP solver HiGHS, as described in~\cite[Section~4]{Bem26}, which takes between 0.412 and 0.499~ms, with an average of 0.432~ms. Note that we use MILP as the GNEP that must be solved at each time step is not monotone, nor do we restrict ourselves to variational solutions; this rules
out most of the available alternative methods for solving linear quadratic GNEPs.}

\begin{table}[t]
\setlength{\tabcolsep}{2.8pt} 
\begin{tabular}{l|r|r|r|r|r|r|r}
\hline
method & unique & $\infty$-many & v-GNE & min norm& welfare & total &CPU time\\ \hline
no split & 12 & 3 & - & -& -& 15 & 6.80~s\\
min-norm & 12 &  -  & - &15 & -& 27 & 12.28~s\\
v-GNE & 12 & 3  & 3 & - & - & 18 & 7.11~s\\ 
welfare & 12 & - & - & - &15&27&12.37~s \\\hline
\end{tabular}
\caption{Critical regions found for the two-mass-spring-damper explicit game-theoretic MPC problem.} \label{tab:two-masses-CRs}
\end{table}

\begin{figure}
    \centering
    \includegraphics[width=\linewidth]{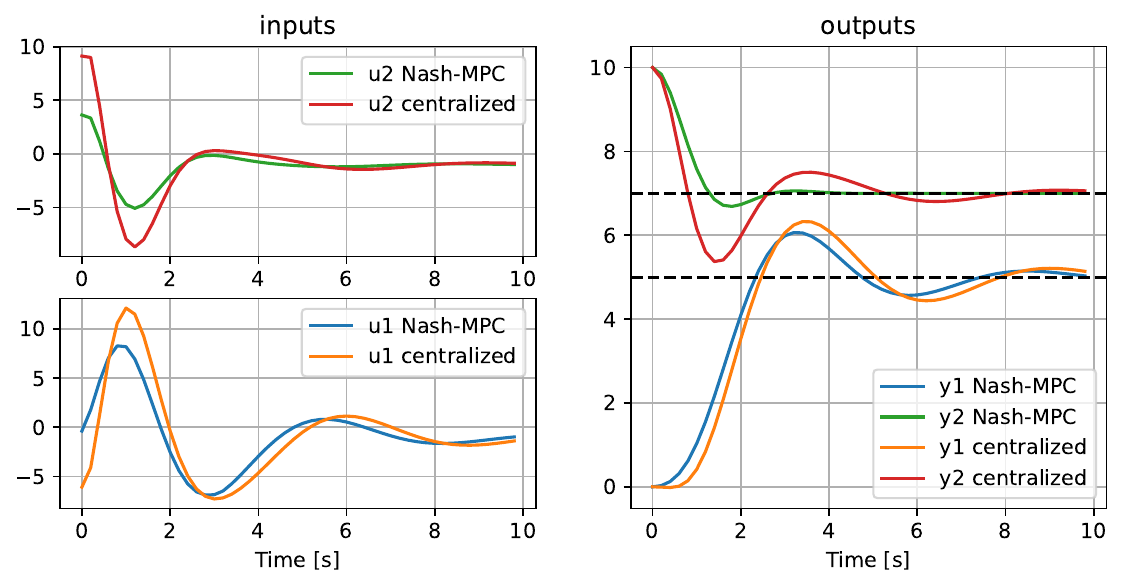}
    \caption{Closed-loop simulation of the two-mass-spring-damper system controlled by two agents using game-theoretic MPC vs centralized MPC}
    \label{fig:nash_mpqp_two_mass_MPC}
\end{figure}

We simulate the closed-loop system for 50 time steps, comparing the \change{explicit game-theoretic MPC controller with the optimization-based MPC.} The results are shown in Figure~\ref{fig:nash_mpqp_two_mass_MPC}, where we can see that both controllers successfully move the masses to their desired positions. However, the closed-loop trajectories are quite different: with \change{game-theoretic MPC}, agent $2$ tracks its reference position more greedily, with agent $1$ adjusting its position error accordingly, while the \change{MPC optimizing the sum of costs} finds a compromise of the agents' objective to track both references. In both cases, the coupling constraint $y_2(t)-y_1(t)\geq\Delta y$ is satisfied at all steps $t$.

Table~\ref{tab:two-masses-CRs} shows the critical regions found for this explicit Nash equilibrium problem. All game-theoretic MPCs yield a very similar closed-loop behavior. Although the following cases have not been recorded in our simulations, in the case of overlapping regions we would have chosen the controller gain associated with the first region that $p$ is found to belong to, and, in case of regions with infinitely many solutions, the value of $y_2$ with smallest norm such that $(p,y_2)$ belongs to the polyhedron in~\eqref{eq:mpQP-CR-partial-rank} (a trivial operation when $y_2\in\rr$.)


\section{Conclusions} \label{sec:Conclusion}
We have presented a methodology and a constructive algorithm to compute explicit solutions to multiparametric GNE problems characterized by convex quadratic agents' cost functions and coupling/local linear inequality constraints. By leveraging existing approaches for solving mpQP problems, we imposed equilibrium conditions of the best responses of each agent with respect to the other agents' actions and parameters of the problem. We also characterized the critical regions in which infinitely many equilibria may occur, offering different criteria to select a specific solution fulfilling a certain desired property, such as minimum-norm solutions and homogeneous Lagrange multipliers for agents to satisfy the coupling constraint (variational GNE). An application of the approach to a battery charging game-theoretic MPC and to a toy mechanical system demonstrated the potential for real-time implementation and interpretability of game-theoretic MPC schemes. Future work will investigate stability properties of explicit GNE solutions and recursive feasibility of the resulting receding horizon implementation. 

\bibliographystyle{IEEEtran}
\bibliography{TAC_explicit_nash_mpc}	

@STRING{IEEETAC = "IEEE Transactions on Automatic Control"}

@article{FB25,
  author          = {F. Fabiani and A. Bemporad},
  title = {An active learning method for solving competitive multi-agent decision-making and control problems},
  journal = IEEETAC,
  volume          = 70,
  number          = 4,
  year            = {2025},
  pages = {2374--2389}
}

@Article{rosen1965existence,
  author     = {J. B. Rosen},
  journal    = {Econometrica},
  title = {Existence and Uniqueness of Equilibrium Points for Concave $N$-Person Games},
  year       = {1965},
  number     = {3},
  pages      = {520--534},
  volume     = {33},
  publisher  = {[Wiley, Econometric Society]},
  url        = {http://www.jstor.org/stable/1911749}
}

@Article{hall2024stability,
  author        = {Hall, Sophie and Liao-McPherson, Dominic and Belgioioso, Giuseppe and Dörfler, Florian},
  title = {Stability Certificates for Receding Horizon Games},
  journal = {arXiv preprint},
  year          = {2024},
  month         = apr,
  doi           = {10.48550/ARXIV.2404.12165},
}

@InCollection{facchinei2009nash,
  author    = {Francisco Facchinei and Jong Pang},
  booktitle = {Convex Optimization in Signal Processing and Communications},
  publisher = {Cambridge University Press},
  title = {{Nash} equilibria: the variational approach},
  year      = {2009},
  chapter   = {12},
  editor    = {Daniel P. Palomar and Yonina C. Eldar},
  month     = {dec},
  pages     = {443--493},
  doi       = {10.1017/cbo9780511804458.013},
  place     = {Cambridge}
}

@Article{facchinei2009generalized,
  author     = {Facchinei, Francisco and Kanzow, Christian},
  journal    = {4OR},
  title = {Generalized {Nash} equilibrium problems},
  year       = {2009},
  number     = {3},
  pages      = {173--210},
  volume     = {5},
  doi        = {10.1007/s10288-007-0054-4}
}

@Article{bemporad2002explicit,
  author  = {A. Bemporad and M. Morari and V. Dua and E.N. Pistikopoulos},
  journal = {Automatica},
  title = {The Explicit Linear Quadratic Regulator for Constrained Systems},
  year    = {2002},
  number  = {1},
  pages   = {3--20},
  volume  = {38}
}

@Article{hall2025limits,
  author        = {Hall, Sophie and Dörfler, Florian and Nax, Heinrich H. and Bolognani, Saverio},
  title = {The Limits of Fairness of the Variational Generalized {Nash} Equilibrium},
  journal = {arXiv preprint},
  year          = {2025},
  month         = apr,
  doi           = {10.48550/ARXIV.2504.03540},
}

@Article{xinyingchen2023guide,
  author = {V. Xinying Chen and J. N. Hooker},
  journal   = {Annals of Operations Research},
  title = {A guide to formulating fairness in an optimization model},
  year      = {2023},
  month     = apr,
  number    = {1},
  pages     = {581--619},
  volume    = {326},
  doi       = {10.1007/s10479-023-05264-y},
  publisher = {Springer Science and Business Media LLC}
}

@Misc{kulkarni2019efficiency,
  author    = {Kulkarni, Ankur A.},
  title = {The Efficiency of Generalized {Nash} and Variational Equilibria},
  year      = {2019},
  doi       = {10.48550/ARXIV.1908.00702},
}

@Article{hall2024receding,
  author    = {Hall, Sophie and Guerrini, Laura and Dörfler, Florian and Liao-McPherson, Dominic},
  journal   = {IFAC-PapersOnLine},
  title = {Receding Horizon Games for Modeling Competitive Supply Chains},
  year      = {2024},
  number    = {18},
  pages     = {8--14},
  volume    = {58},
  doi       = {10.1016/j.ifacol.2024.09.002},
  publisher = {Elsevier BV}
}

@Article{atzeni2012demand,
  author     = {Atzeni, Italo and Ord{\'o}{\~n}ez, Luis G and Scutari, Gesualdo and Palomar, Daniel P and Fonollosa, Javier Rodr{\'\i}guez},
  journal    = {IEEE Trans. Smart Grid},
  title = {Demand-side management via distributed energy generation and storage optimization},
  year       = {2012},
  month      = jun,
  number     = {2},
  pages      = {866--876},
  volume     = {4},
  doi        = {10.1109/tsg.2012.2206060},
  publisher  = {Institute of Electrical and Electronics Engineers ({IEEE})}
}

@InProceedings{hall2022receding,
  author    = {Hall, Sophie and Belgioioso, Giuseppe and Liao-McPherson, Dominic and Dorfler, Florian},
  booktitle = {2022 IEEE 61st Conference on Decision and Control (CDC)},
  title = {Receding Horizon Games with Coupling Constraints for Demand-Side Management},
  year      = {2022},
  pages = {3795--3800},
  doi       = {10.1109/CDC51059.2022.9992497}
}

@InBook{bassanini2002allocation,
  author    = {Bassanini, A. and La Bella, A. and Nastasi, A.},
  pages     = {1--17},
  publisher = {Springer US},
  title = {Allocation of Railroad Capacity Under Competition: A Game Theoretic Approach to Track time Pricing},
  year      = {2002},
  booktitle = {Transportation and Network Analysis: Current Trends},
  doi       = {10.1007/978-1-4757-6871-8_1}
}

@Article{lecleach2022algames,
  author    = {Le Cleac’h, Simon and Schwager, Mac and Manchester, Zachary},
  journal   = {Autonomous Robots},
  title = {{ALGAMES}: a fast augmented {Lagrangian} solver for constrained dynamic games},
  year      = {2022},
  number    = {1},
  pages     = {201--215},
  volume    = {46},
  doi       = {10.1007/s10514-021-10024-7},
  publisher = {Springer}
}

@Book{pavel2012game,
  author    = {Pavel, Lacra},
  publisher = {Birkhäuser Boston},
  title = {Game Theory for Control of Optical Networks},
  year      = {2012},
  doi       = {10.1007/978-0-8176-8322-1}
}

@PhdThesis{dreves2011globally,
  author = {Axel Dreves},
  school = {Universit{\"a}t W{\"u}rzburg},
  title = {Globally Convergent Algorithms for the Solution of Generalized {Nash} Equilibrium Problems},
  year   = {2011},
  type   = {doctoralthesis}
}

@Book{facchinei2003finite,
  author    = {Francisco Facchinei and Jong-Shi Pang},
  publisher = {Springer-Verlag New York Inc.},
  title = {Finite-Dimensional Variational Inequalities and Complementarity Problems},
  year      = {2003},
  month     = feb,
  volume    = {II}
}

@Article{facchinei2007generalized,
  author    = {Francisco Facchinei and Andreas Fischer and Veronica Piccialli},
  journal   = {Mathematical Programming},
  title = {Generalized {Nash} equilibrium problems and {Newton} methods},
  year      = {2007},
  month     = jul,
  number    = {1-2},
  pages     = {163--194},
  volume    = {117},
  doi       = {10.1007/s10107-007-0160-2},
  publisher = {Springer Science and Business Media LLC}
}

@Article{pang2005quasi,
  author    = {Pang, Jong-Shi and Fukushima, Masao},
  journal   = {Computational Management Science},
  title = {Quasi-variational inequalities, generalized {Nash} equilibria, and multi-leader-follower games},
  year      = {2005},
  month     = jan,
  number    = {1},
  pages     = {21--56},
  volume    = {2},
  doi       = {10.1007/s10287-004-0010-0},
  publisher = {Springer Science and Business Media LLC}
}

@Article{belgioioso2022distributed,
  author       = {Belgioioso, Giuseppe and Yi, Peng and Grammatico, Sergio and Pavel, Lacra},
  journal      = {IEEE Control Systems Magazine},
  title = {Distributed generalized {Nash} equilibrium seeking: An operator-theoretic perspective},
  year         = {2022},
  month        = {aug},
  number       = {4},
  pages        = {87--102},
  volume       = {42},
  doi          = {10.1109/MCS.2022.3171480},
  publisher    = {IEEE}
}

@Article{yi2019operator,
  author    = {Peng Yi and Lacra Pavel},
  journal   = {Automatica},
  title = {An operator splitting approach for distributed generalized {Nash} equilibria computation},
  year      = {2019},
  month     = apr,
  pages     = {111--121},
  volume    = {102},
  doi       = {10.1016/j.automatica.2019.01.008},
  publisher = {Elsevier BV}
}

@Article{bianchi2022fast,
  author    = {Mattia Bianchi and Giuseppe Belgioioso and Sergio Grammatico},
  journal   = {Automatica},
  title = {Fast generalized {Nash} equilibrium seeking under partial-decision information},
  year      = {2022},
  month     = feb,
  pages     = {110080},
  volume    = {136},
  doi       = {10.1016/j.automatica.2021.110080},
  publisher = {Elsevier BV}
}

@Article{belgioioso2023semi,
  author    = {Belgioioso, Giuseppe and Grammatico, Sergio},
  journal   = {IEEE Transactions on Automatic Control},
  title = {Semi-decentralized generalized {Nash} equilibrium seeking in monotone aggregative games},
  year      = {2023},
  number    = {1},
  pages = {140--155},
  volume    = {68},
  doi       = {10.1109/TAC.2021.3135360},
  publisher = {IEEE}
}

@InProceedings{arnstroem2024high,
  author       = {Arnstr{\"o}m, Daniel and Axehill, Daniel},
  booktitle = {2024 IEEE 63rd Conference on Decision and Control (CDC)},
  title = {A high-performant multi-parametric quadratic programming solver},
  year         = {2024},
  organization = {IEEE},
  pages        = {303--308}
}

@Article{schaller2025automatic,
  author        = {Schaller, Maximilian and Arnström, Daniel and Bemporad, Alberto and Boyd, Stephen},
  title = {Automatic Generation of Explicit Quadratic Programming Solvers},
  journal = {arXiv preprint},
  year          = {2025},
  month         = jun,
  doi           = {10.48550/ARXIV.2506.11513},
}

@article{MR64,
    author="O.L. Mangasarian and J.B. Rosen",
    title="Inequalities for stochastic nonlinear programming problems",
    journal="Operations Research",
    year="1964",
    volume="12",
    pages="143--154"
}

@book{Fia83,
  author =  {A.V. Fiacco},
  title = {Introduction to sensitivity and stability analysis in nonlinear programming},
  year =    {1983},
  publisher={Academic Press},
  address = {London, U.K.}
}

@article{GBN11,
    title = {A novel approach to multiparametric quadratic programming},
    author={A. Gupta and S. Bhartiya and P.S.V. Nataraj},
    journal={Automatica},
    volume={47},
    number={9},
    pages={2112--2117},
    year={2011}
}

@article{tondel2003algorithm,
	title = {An algorithm for multi-parametric quadratic programming and explicit {MPC} solutions},
	author={T{\o}ndel, P. and Johansen, T. and Bemporad, A.},
	journal={Automatica},
	volume={39},
	number={3},
	pages={489--497},
	year={2003},
	publisher={Elsevier}
}

@book{BV04,
  author =       {S. Boyd and L. Vandenberghe},
  title = {Convex Optimization},
  publisher = {Cambridge University Press},
  address={New York, NY, USA},
  year =         {2004}
}

@article{BBM00mplp,
  author =  {F. Borrelli and A. Bemporad and M. Morari},
  title = {A Geometric Algorithm for Multi-Parametric Linear Programming},
  year =    {2003},
  journal = {Journal of Optimization Theory and Applications},
  volume={118},
  number={3},
  pages={515--540},
  month=sep
}

@article{DP99b,
  author =  {V. Dua and E.N. Pistikopoulos},
  title = {An algorithm for the solution of multiparametric mixed integer
             linear programming problems},
  journal = {Annals of Operations Research},
  year =    {2000},
  volume={1},
  month=dec,
  pages={123--139}
}

@Article{liu2024input,
  author        = {Liu, Mushuang and Kolmanovsky, Ilya},
  title = {Input-to-State Stability of {Newton} Methods in {Nash} Equilibrium Problems with Applications to Game-Theoretic Model Predictive Control},
  journal = {arXiv preprint},
  year          = {2024},
  month         = dec,
  doi           = {10.48550/ARXIV.2412.06186},
}

@article{Bem26,
    author={A. Bemporad},
    title={{NashOpt}: A {Python} Library for Computing Generalized {Nash} Equilibria and Game Design},
    note = {code available at \url{https://github.com/bemporad/nashopt}},
    year=2026,
    journal = {Optimization Methods \& Software},
}


\end{document}